# Computer Vision Methods for Automating Turbot Fish Cutting

Fernando Martín-Rodríguez, Fernando Isasi-de-Vicente, Mónica Fernández-Barciela.
fmartin@tsc.uvigo.es, fisasi@tsc.uvigo.es, monica@tsc.uvigo.es.
atlanTTic research center for Telecommunication Technologies, University of Vigo,
Campus Lagoas Marcosende S/N, 36310 Vigo, Spain.

*Abstract*- **This paper is about the design of an automated machine to cut turbot fish specimens. Machine vision is a key part of this project as it is used to compute a cutting curve for specimen's head. This task is impossible to be carried out by mechanical means. Machine vision is used to detect head boundary and a robot is used to cut the head. Binarization and mathematical morphology are used to detect fish boundary and this boundary is subsequently analyzed (Hough transform/convex hull) to detect key points and thus defining the cutting curve. Afterwards, mechanical systems are used to slice fish to get an easy presentation for end consumer (as fish fillets than can be easily marketed and consumed).**

*Keywords*- **turbot fish, food industry, machine vision, mathematical morphology, convex hull, Hough transform.**

## I. Introduction

There is some interest in food industry to be able to automatically slice turbot fishes to get a new commercial presentation. It would consist of fish fillets with no fish bones, very easy to cook and to eat, probably very attractive to customers. Turbots are flat fishes; they have their two eyes on the same side as they always swim on the bottom. They also have a very delicate flavor and high commercial interest. Nowadays, they are farmed in numerous places, so that it is easy and cheap getting big amounts of individuals with great size variety.

Its special shape makes it difficult to create an automatic cutter. The main problem is cutting the fish head, as it is necessary to develop a cutting curve that will depend on specimen size. Formerly, this cutting was made by specialized personnel but it is an unpleasant and dangerous activity because specimens are very slippery and can cause that workers hurt themselves with their own knives (they use very big knives).

That's the reason for using computer vision as a means of automatically detecting the necessary cutting curve and instructing a robot for cutting the head away (XML will be used to communicate the curve to an industrial robot). Afterwards, purely mechanical systems are used to slice fish to get an easy presentation for end consumer.

Due to the extremely complex texture of fish skin where computing gradients would yield extremely noisy results, we will base ourselves on analyzing contour curves instead of trying to detect head contour over fish body.

We have not found journal references for this particular problem.

## II. Materials and Methods

### A. Image capturing

The slicing machine is designed to place each specimen on a proper, white colored, surface (specimens are of dark color). Camera position will be zenithal so that we get a whole image of the flat fish. We will use LED illumination, very probably LED bars used for indirect lighting (specimen and camera are inside a steel box with reflective walls). We will use a matrix (field) camera like JAI GO Series [1].

At this moment, we have developed computer vision algorithms using color images that were captured manually at a fish processing plant (figure 1). Camera for test images was a commercial DSLR (Canon EOS 1100D).

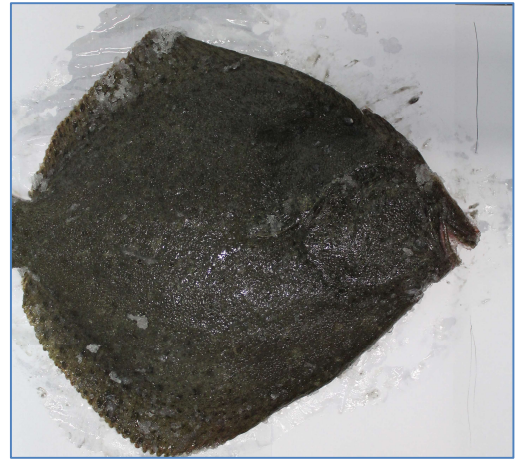

Fig. 1. Example test image.

### B. Preprocessing

Image preprocessing will consist of segmenting the specimen. We start by using the well-known Otsu threshold [2]. This method computes optimum threshold for a bimodal image (histogram has only two dominant peaks) using conditional variances. We compute Otsu threshold and we call it $u_0$. Nevertheless, we will not binarize image immediately. Conversely, we apply a gamma correction ($Im[x,y]^{gamma}$) that makes brighter dark objects that do not correspond to fish (mainly the stains on the white background surface). Gamma is computed as $log(max)/log(u0)$. Making

max=0.80 we ensure that the histogram central minimum (Otsu threshold) goes high in luminance. Applying Otsu again and binarizing the darkest object (fish) is the only one selected under threshold.

After binarization, image is cleaned removing small objects that are not part of the specimen. This is achieved using mathematical morphology [3]. To assure that structuring element sizes are appropriate, images are scaled into a fixed size of 2000 lines. Now, we describe the morphological filters used, step by step.

1. Opening with a circle of radius equal to 20 <u>Purpose</u>: erasing small objects. We use "open by reconstruction" for best results.
2. Closing with a circle of radius 10. <u>Purpose</u>: filling holes inside the fish mask. "Closing by reconstruction" is used again to yield cleaner results.
3. Labeling and extracting the biggest connected object.

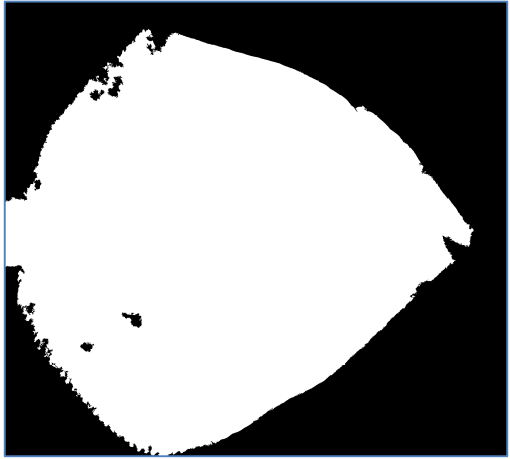

Fig. 2. Example test image.

Final result is shown in figure 2. Remaining holes are not important as we will focus on fish contour. Using the former image, it is easy to obtain the region of interest (ROI), id EST: the front part including head. Nose is detected as the rightmost active (white) point.

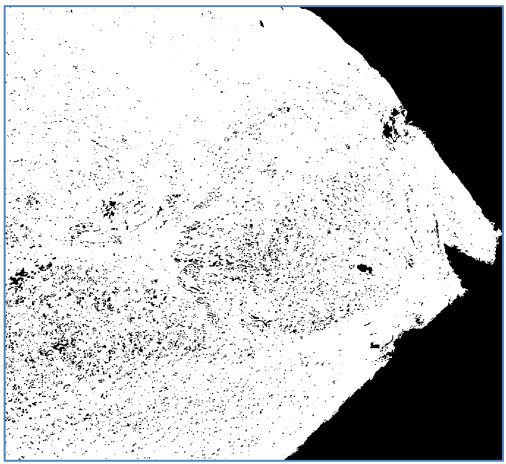

Fig. 3. ROI: Region of interest.

## C. Critical points computation

For computing a cutting curve, we base ourselves on analyzing the contour of the defined ROI (figure3). Note that ROI is obtained with coordinates computed from figure 2, but image data are taken from original image (figure 1) and then binarized. Two concave discontinuities in this contour can define the starting and finishing points of our desired cutting curve.

Two methods were tested for detecting these points: one based on Hough transform and other based on convex hull.

C.1.- Hough transform application

Hough transform [4] is able to detect straight lines. In this case, we have two dominant straight lines. Interest points should be detected as prominent errors of this linear approximation.

First, we define and compute the specimen contour as:
<<**The curve that is made with the last (rightmost) active (white) point on each mask line.**>>

As we said before, using Hough transform, we can find the two dominant straight lines. One descending from the upper part until the nose and the one that ascends from bottom to the same point (nose). See figure 4. With the equations of these two lines, we will find the critical points in the contour. More precisely, we compute distance between each point in the contour and the nearest straight line (from the two referred ones).

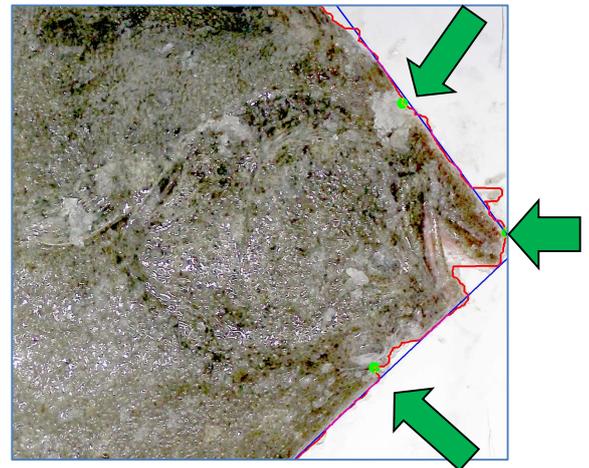

Fig. 4. Straight lines and critical points detected.

That distance is processed as a signal, so that we can detect (figure 4):

- Beginning of the head (upper arrow). It is defined by the first local maximum of distance function. In fact, it is the local maximum that is nearest to the begin or to the end of such signal. Detecting this point allows us to know in what position we have specimen's eyes. Laying in this manner (eyes up and nose looking right) it is more frequent that the eye that is closest to the contour is in the lower position, but some specimens have it reversed. For maximum detection, we have used "neighborhood peak detection" [5] that is based on studying the dominance of each relative maximum to its neighborhood.
- Once we have detected the first critical point, we analyze the other half curve beyond the central point (or above for "misplaced-eyes" specimens). We do not consider the neighborhood of the nose and we search for the first local minimum (using the same method in [5]). This will allow us to detect head ending which coincides with an eye (lower arrow).

The third critical point is the nose that was already known from the preprocessing stage (it is the rightmost point in the contour). In this last image, we have applied a gamma correction that allows distinguishing better the color texture in the fish body, while "burning" to the extreme the white background. Nevertheless, detecting head contour using textures seems very difficult.

Figure 5 illustrates critical point detection: horizontal axis is simply labeled with the sample count, the central value corresponds approximately to the nose; vertical axis measures distance from contour points to the straight lines, units are simply pixels.

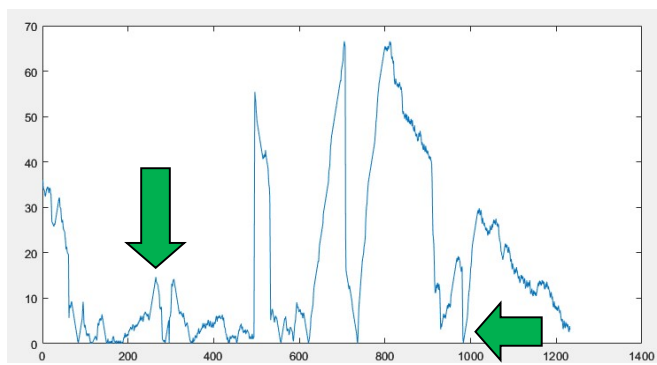

Fig. 5. Critical points computation.

C.2.- Convex hull application

In figure 6, we represent the convex hull of ROI (figure 3). We compute the difference between both contours, id EST: we find rightmost active point on each line for both images and we subtract X-coordinates. This yields a one-dimensional error signal (figure 7) that can be processed to detect critical points.

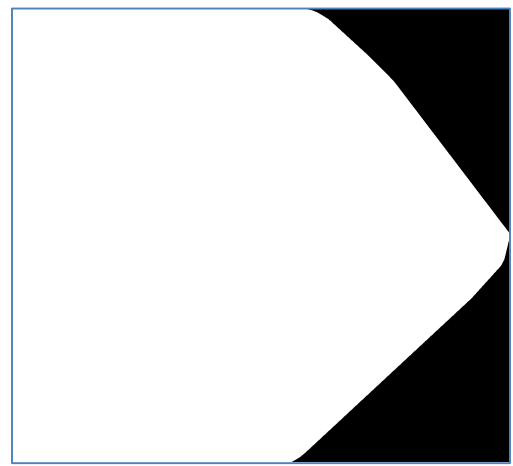

Fig. 6. Example test image.

Horizontal axis in figure 7 is again labeled with the sample count (or image line number), central value corresponds approximately to the nose. Vertical axis measures distance from contour points to convex hull, units are pixels.

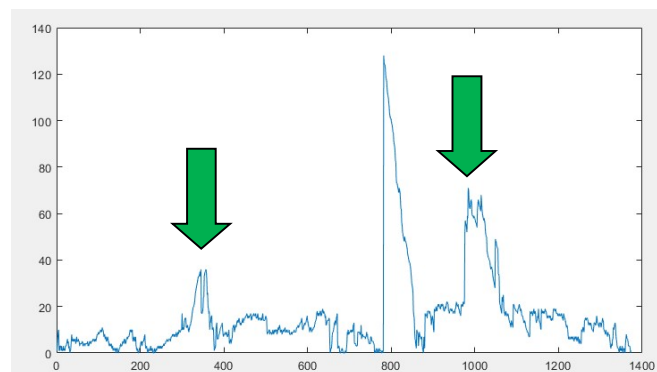

Fig. 7. Example test image.

Critical point detection is as follows: error signal is divided into two regions, before and after the specimen nose (central point and, usually, absolute maximum). These two regions are studied, detecting:

- Head beginning (left green arrow): it is defined by the maximum of the first region.
- Head ending (which coincides with an eye, right arrow): it will be the maximum of the second region.

In figure 8, we show more graphically the result of point detection. Almost always, the "eye that touches the contour" is in the lower side. This is true for more than 90% of specimens. There are some special cases where this eye is in the upper side. Notice that with our method, these special cases will be treated properly.

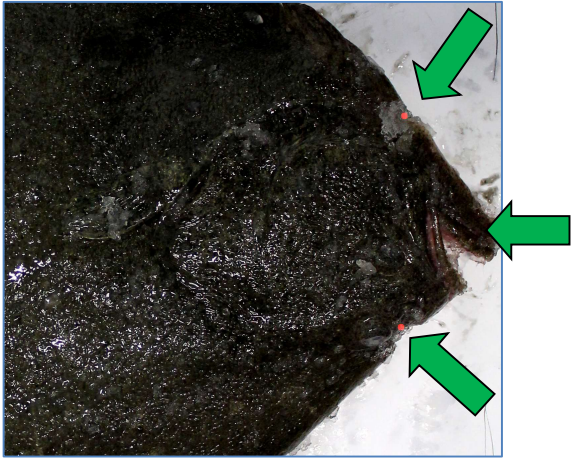

Fig. 8. Detected critical points: head beginning, nose, head ending.

*D. Cutting curve computation*

To compute the cutting curve, we tried first a parabola passing by the two head endings. Afterwards, we changed our minds and decided to use an ellipse.

D.1.- Parabola

The parabola should pass by the two head endings: upper and lower critical points, computed just now. As it is a second order equation, we need three points to solve for the polynomial coefficients. Third point was heuristically determined from the three points draw in figure 8.

The third point will be "approximately" the vertex of the parabola. It will be at the same height than nose and at a horizontal distance of it, equal to three times d. Let d be the maximum between the two horizontal distances with head beginning and ending points. Calculation is illustrated in figure 9, result is shown figure 10.

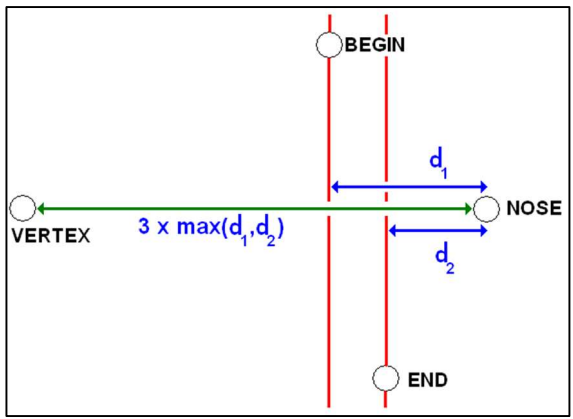

Fig. 9. Parameters for parabola computation.

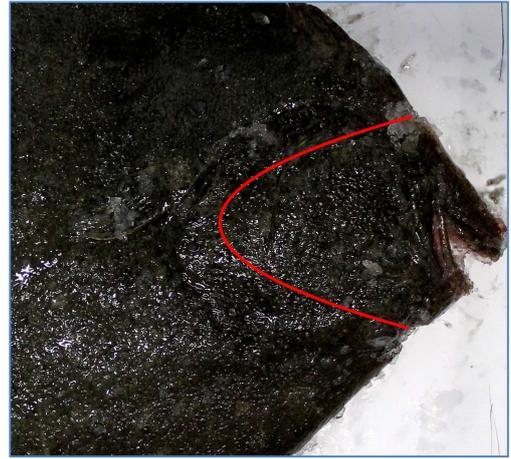

Fig. 10. Cutting curve: parabola.

D.2.- Ellipse

Our ellipse will have as minor (vertical) axis, the line between the two endings. Major axis will be twice the minor axis (eccentricity equal to $\sqrt{3}/2$).

In this method, we force the two endings to be aligned vertically. We consider as true the point with maximum X coordinate (rightmost position). If necessary, we substitute the second point with the intersection of specimen boundary with the vertical line passing by the first point.

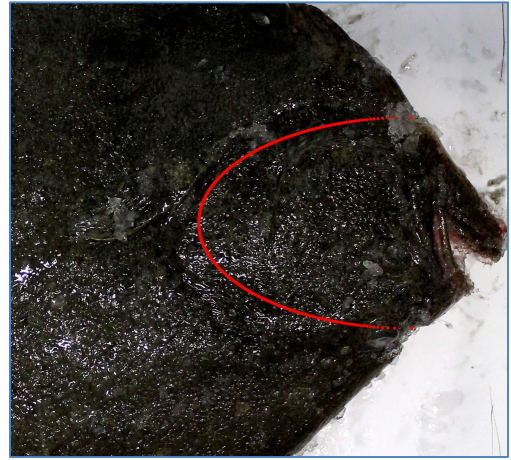

Fig. 11. Cutting curve: ellipse.

III. RESULTS AND DISCUSSION

Best results (more accurate cutting curves) were found for the convex hull method. Trying with fifty images, 100% of them got a valid cutting curve.

Considering the two options for cutting curve, ellipse is normally the preferred one. Anyway, these options can be set as configurable in the final machine so that cutting can be tuned.

*A. Conclusion and future lines*

We have found an easy yet effective method for computing automatically the cutting curve for turbot fish head. This is the solution for a previously unsolved problem.

Main future line is the integration with a final machine that, at the time of writing, still does not exist. Cutting curve must be sent from an industrial computer that runs the vision process to a robotic arm that will cut the head. This will be done using a XML file with a set of 2D points defined via (x,y) coordinates in millimeters.

ACKNOWLEDGEMENT

This work was partially funded by the "Centro para el Desarrollo Tecnológico Industrial" (CDTI, www.cdti.es), project: CO-0170-11.